\documentclass[epj,twocolumn]{webofc}
\usepackage[varg]{txfonts} 

\usepackage{amsmath}
	  \usepackage{amssymb}
	 \usepackage{graphics}
	  \usepackage{epsfig}
	  \usepackage[dvipdfm]{hyperref}
	  
\DeclareMathOperator{\Df}{{\mathcal{ D}}}      
\DeclareMathOperator{\lagr}{\mathcal{ L}} 

\woctitle{Quarks 2016}

\begin{document}

\title{Applicability of the Wigner functional approach to  evolution of quantum fields }

\author{\firstname{Andrey} \lastname{Leonidov}\inst{1,2,3}\fnsep\thanks{\email{leonidov@lpi.ru}} \and
        \firstname{Anna} \lastname{Radovskaya}\inst{1}\fnsep\thanks{\email{raan@lpi.ru}}
}

\institute{P.N. Lebedev Physics Institute of the Russian Academy of Sciences, Moscow, Russia 
\and
           Moscow Engineering Physics Institute, Moscow, Russia
\and    
           Moscow Institute of Physics and Technology, Moscow, Russia
          }

\abstract{Evolution of highly excited quantum   field is considered in the framework of Keldysh formalism .
It is demonstrated that leading order (LO) term of semiclassical approximation appears as well-known Classical Statistical Approximation (CSA).
In simple case of spatially homogeneous scalar field analytical expressions for leading and next-to-leading (NLO) order are presented. 
It is shown that the range of applicability of 
CSA strongly depends on the properties of the initial state of the system.
}
\maketitle
\section{Introduction}\label{intro}
The problem of description  of  the evolution of highly excited quantum  fields has recently attracted considerable 
attention of theorists working on problems in various domains of theoretical physics. 
The relevant examples include, in particular, studies of initial stages of ultrarelativistic heavy ion collisions \cite{initial1, kurkela3}, heating of the Universe after
inflation \cite{cosm_1, cosm_3}, particle production in ultrastrong fields \cite{emil_1, gelis}, thermalization in  ultracold quantum gases
 \cite{berges_gases, Lee}.
All the above-mentioned systems share the basic common feature  -  their initial state can not be described as the vacuum one. 
This means that the standard S-matrix formalism is not applicable here.  In fact, for highly excited fields 
it is necessary to give a correct specification of the initial state and be able to describe its subsequent temporal evolution.  
The Keldysh technique \cite{Keldysh, berges_review_2} provides all the necessary technical tools for solving such problems. 

In the previous work \cite{LR1} we have demonstrated that the Keldysh formalism applied to description of evolution of 
excited fields results in its Leading Order in  so-called Classical Statistical Approximation (CSA). 
CSA  is based on the fact that summation of leading quantum corrections
 can be cast in the form of integration over initial conditions for classical trajectories 
 with the weight given by the Wigner functional. It is interesting to note that similar approach was used in atomic and chemical physics 
 \cite{chem1},\cite{chem2}.
 
 In this paper we analyze the applicability of the Wigner functional approach
 to evolution of highly excited quantum fields by calculating the Next-to-Leading corrections to several observables.
 We show that this applicability does directly depend on the properties of the initial state of the system and find out range of the validity 
 of the Wigner functional approach  considering  homogeneous scalar field model as an example.

\section{General formalism} 
In this section we will shortly remind  general formalism for
calculation of observables within the Keldysh technique  suitable to our problem. 
Out of equilibrium an expectation value of observable $F(\hat\varphi)$ at the moment $t_1$ 
can be calculated  as a trace with density matrix as
\begin{multline}
  \langle F(\hat\varphi) \rangle _{t_1}=tr (F(\hat\varphi) \hat\rho(t_1))\\
    =\int \mathfrak{ D}\xi(\vec x)\  F(\xi) \langle \xi|\hat U(t_1,t_0) \hat\rho(t_0) \hat U(t_0,t_1)|\xi \rangle,
%  = \int \mathfrak{ D}\xi \int \mathfrak{ D}\xi_1 \int \mathfrak{ D}\xi_2\  F(\xi) \langle\xi|\hat U(t_1,t_0)|\xi_1\rangle
%  \langle\xi_1|\hat\rho(t_0)|\xi_2\rangle
%\langle\xi_2| \hat U(t_0,t_1)|\xi\rangle.\label{genev}
\end{multline}
where evolution of the density matrix $\hat\rho(t)$ is governed by the evolution operator $\hat U(t,t_0)$ 
\begin{equation}
 \hat\rho(t)=\hat U(t,t_0) \hat\rho(t_0) \hat U(t_0,t),
\end{equation}
  $|\xi\rangle$ is an eigenstate of the field operator 
  \mbox{$\hat \varphi(\vec x) |\xi\rangle = \xi(\vec x)|\xi\rangle$}   and 
 $\int \mathfrak{D}\xi(\vec x)$ is a path integral over all possible 3-d functions originating from
 unity operator \mbox{$\hat 1 =\int \mathfrak{D}\xi(\vec x)\ |\xi\rangle \langle\xi|$.}
 
   The matrix elements of the evolution operator can be represented as 
   the path integrals over 4-d functions $\mathcal{D}\eta(t,\vec x)$
    \begin{gather*}
  \langle \xi|\hat U(t_1,t_0)|\xi_1 \rangle
    = \int\limits_{\eta_F(t_0,\vec x)=\xi_1(\vec x)}^{\eta_F(t_1,\vec x)
  =\xi(\vec x)} \Df \eta_F(t,\vec x) e^{i S[\eta_F]},\\
\langle\xi_2|\hat U(t_0,t_1)|\xi\rangle = \int\limits_{\eta_B(t_0,\vec x)=\xi_2(\vec x)}^{\eta_B(t_1,\vec x)
 =\xi(\vec x)} \Df \eta_B(t,\vec x) e^{-i S[\eta_B]}
  \end{gather*}
 
Here $\eta_F(t,\vec x)$  and $\eta_B(t,\vec x)$  are the fields that lie on the  forward ($\eta_F$)
and backward ($\eta_B$) sides of Keldysh contour (see \cite{LR1} for details). Combining all together we obtain 
\begin{gather}
   \langle F(\hat\varphi) \rangle _{t_1}
  = \int \mathfrak{ D}\xi \int \mathfrak{ D}\xi_1 \int \mathfrak{ D}\xi_2\ \langle\xi_1|\hat\rho(t_0)|\xi_2\rangle\ \times\\
    F(\xi) \int\limits_{\eta_F(t_0,\vec x)=\xi_1(\vec x)}^{\eta_F(t_1,\vec x)
  =\xi(\vec x)} \Df \eta_F(t,\vec x) \int\limits_{\eta_B(t_0,\vec x)=\xi_2(\vec x)}^{\eta_B(t_1,\vec x)
 =\xi(\vec x)} \Df \eta_B(t,\vec x)\ e^{iS[\eta_F]-iS[\eta_B]}.\nonumber
\end{gather}

The next step is to perform technical trick and extend Keldesh contour to infinity. After that we can introduce new fields  as
\begin{equation}
\phi_c = \frac{\eta_F+\eta_B}{2}, \qquad \phi_q = \eta_F -\eta_B 
\end{equation}  
and define the Keldysh action  $ S_K[\eta_F,\eta_B] = S[\eta_F] - S[\eta_B]$ in order to obtain general expression
\begin{gather} \label{general2}
    \langle F(\hat\varphi)\rangle_{t_1}
   = \int \mathfrak{D}\chi_1 \int \mathfrak{D}\xi_1 \int \mathfrak{D}\xi_2\ \langle\xi_1|\hat\rho(t_0)|\xi_2\rangle\ \times  \\
   \int\limits_{\phi_c(t_0,\vec x)=\frac{\xi_1(\vec x)+\xi_2(\vec x)}{2}}^{\phi_c(\infty,\vec x) =\chi_1(\vec x)}
  \Df \phi_c
  \int\limits_{\phi_q(t_0,\vec x)=\xi_1(\vec x)-\xi_2(\vec x)}^{\phi_q(\infty,\vec x)=0}  \Df \phi_q
  \ F(\phi_c(t_1))
  \ e^{i S_K[\phi_c,\phi_q]} .\nonumber
\end{gather}

  Eq.(\ref{general2}) represents our main tool for calculation of observables 
  however we need to specify Lagrangian in order to calculate certain results.
  
  \section{Scalar $\varphi^4$ theory} 
  The simplest model which has necessary nontrivial dynamical behavior is a scalar $\varphi^4$ model with quartic interaction term.
  \begin{gather}\label{static_lagr}
  \lagr = \frac{1}{2}\partial_{\mu}\varphi\partial^{\mu}\varphi - \frac{g^2}{4} \varphi^4 + J \varphi,\\
  S = \int d^4x \lagr. \nonumber
  \end{gather}
 Here $J(t,\vec x)$ is an auxiliary source which we keep  in order to perform semiclassical decomposition.

For the Lagrangian (\ref{static_lagr})  the Keldysh actions reads
\begin{gather}
 S_K[\phi_c,\phi_q] =
 \int d^3x\ \dot\phi_c(t_0,\vec x)(\xi_1(\vec x)-\xi_2(\vec x))\nonumber\\
 - 
 \int d^3x \int\limits_{t_0}^{\infty}dt\ \Big(\phi_q A[\phi_c]
  -\frac{g^2 }{4}\phi_c\phi_q^3\Big),\nonumber \\
 A[\phi_c]=[\partial_{\mu}\partial^{\mu}\phi_c+g^2\phi_c^3 - J].
 \label{static_action}
\end{gather}
Here term $\phi_q(t_0,\vec x) = \xi_1(\vec x) - \xi_2(\vec x)$ appears from integration by parts. 
Note that $A[\phi_c]=0$ corresponds to projecting onto the classical equation of motion for the Lagrangian Eq.(\ref{static_lagr}).
It is important that the source $J$ connected to both fields on upper and lower sides of Keldysh contour in similar way so we get term 
$J (\eta_F - \eta_B)= J \phi_q$. 

It is clear that path integrals of (\ref{general2}) with this action  can not be calculated explicitly.
The systematic procedure we employ is expansion in $\phi_q$ in (\ref{static_action}) around its saddle-point value. 
This expansion is, in fact, a semiclassical one. This can be seen by restoring $\hbar$ in the action and replacing \mbox{$\phi_q \to \hbar \phi_q$}
so the only remaining dependence on $\hbar$  is in $\phi_q^3$ term which is proportional to $\hbar^2$ and  
\begin{gather}
e^{ -i\frac{g^2 \hbar^2}{4}\int\limits_{t_0}^{\infty}dt \int d^3 x\ \phi_c\phi_q^3}= \underbrace{1}_{LO}
 -\underbrace{\frac{ig^2\hbar^2 }{4}\int\limits_{t_0}^{\infty}dt \int d^3 x\ \phi_c\phi_q^3}_{NLO}+....
 \label{expansion}
 \end{gather}
This expansion shows that the LO contribution contains quantum fluctuation up to one loop order. 

 \subsection{Leading Order}
The Leading Order (LO) contribution to observables corresponds to the first term in decomposition (\ref{expansion}). 
In this case we are able to perform functional integration over field $\phi_q$ resulting with the functional delta-function which project $\phi_{c}$ 
onto the solution of classical equation of motion of the Lagrangian (\ref{static_lagr}). As the next step we insert the "initial velocity"
as $1=\int\mathfrak{D}p(\vec x)
  \delta(\tilde p(\vec x)-\dot\phi_c(t_0,\vec x))$ in order to integrate over $\phi_{c}$ (see \cite{LR1} for details). After
  substitution $\alpha = \frac{\xi_1+\xi_2}{2},\ \beta = \xi_1 - \xi_2$ we end up with 
 \begin{multline} \label{static_LO}
  \langle F(\hat\varphi)\rangle_{t_1}=\\
   = \int \mathfrak{D}\alpha(\vec x)  \mathfrak{D} p(\vec x) f_W[\alpha(\vec x),p(\vec x)),t_0] F(\phi_{cl}(t_1,\vec x)),
  \end{multline}
  where the Wigner functional is commonly defined as
  \begin{multline}
  f_W[\alpha(\vec x),p(\vec x),t_0] =\int \mathfrak{D}\beta(\vec x) 
   \Big\langle\alpha+\frac{\beta}{2}\Big|\hat\rho(t_0)\Big|\alpha-\frac{\beta}{2}\Big\rangle \\
   \times \exp\Bigg(i \int  d^3 x\ p(\vec x) \beta(\vec x)  \Bigg).
   \end{multline}
   Here $\phi_{cl}$ is the solution of classical equation of motion 
    \begin{gather}
   \partial_{\mu}\partial^{\mu}\phi_{cl}+g^2\phi_{cl}^3 = 0
   \end{gather}
   with initial conditions given by
   \begin{gather}
   \phi_{cl}(t_0,\vec x) = \alpha(\vec x), \quad
   \dot\phi_{cl}(t_0,\vec x) = p(\vec x)
 \end{gather}
 and at zero axillary source $J(t,\vec x)$.

Expression  (\ref{static_LO}) was earlier derived by other methods in \cite{initial0,initial1},\cite{MM}
see also \cite{berges_review_1},\cite{Jeon}. The general  recipe for calculation at the LO level is the following:
in order to obtain expectation value of observable at a given time moment $t_1$ one need to find 
solution of classical equation of motion as a function of the field and its time derivative at the initial time
and average over initial conditions with weight of the Wigner functional.  Note that choice of the Wigner functional (or density matrix at initial time)
lies outside the method itself
and depends on problem that should be solved. 
For certain choice of the Wigner functional the expression (\ref{static_LO}) represent the 
Classical Statistical Approximation widely used in the context of ultrarelativistic heavy ion collisions.

Let us introduce new notation for averaging over initial conditions 
\begin{gather}\label{ic}
 \langle \mathcal{O}\rangle_{i.c.} 
   = \int \mathfrak{D}\alpha(\vec x)  \mathfrak{D} p(\vec x)  f_W[\alpha(\vec x),p(\vec x)),t_0]\ \mathcal{O}.
\end{gather}
Then we can rewrite (\ref{static_LO}) simply as 
\begin{gather}
 \langle F(\hat\varphi)\rangle_{t_1}^{LO} = \langle F(\phi_{cl}(t_1,\vec x))  \rangle_{i.c.} .
\end{gather}

 \subsection{Next-to-Leading Order}
In order to decide if  expansion over $\phi_q$ is valid we should consider Next-to-Leading Order (NLO) term in (\ref{expansion}).
With this $\phi_q^3$ part the path integration over $\phi_q$ can not be done as easy as at LO level. However one can note that each 
$\phi_q$ can be replaced by functional derivative over source $J$ due to $\phi_q J $ term in the Keldysh action (\ref{static_action})
\begin{gather}
 \frac{\delta }{\delta J(t,\vec x)} e^{i S_K[\phi_c,\phi_q]} = -i \phi_q(t,\vec x)e^{i S_K[\phi_c,\phi_q]}.
\end{gather}
Then  we can perform functional integration over $\phi_q$ and $\phi_c$ in order to obtain  answer for expectation value of the observable 
up to NLO level
\begin{gather}\label{static_NLO}
 \langle F(\hat\varphi)\rangle_{t_1}^{LO+NLO} = 
  \Bigg\langle F(\phi_{cl}(t_1,\vec x))\\
  + \frac{g^2}{4} \int\limits_{t_0}^{t_1}dt_2 \int d^3 x_2
  \phi_{cl}(t_2,\vec x_2)\frac{\delta^3 F(\phi_{cl}(t_1,\vec x))}{\delta J^3(t_2,\vec x_2)}\Bigg|_{J=0}  \Bigg\rangle_{i.c.}.\nonumber
\end{gather}
The expression above shows that there is no necessity in any new information for evaluation of NLO correction.
We should find classical trajectory as a function of initial conditions, perform three variations over auxiliary source, integrate over  
intermediate time and average with the Wigner functional. These calculations seem to be rather tedious however there is a technical trick.

Let us define $k$-th variation of the classical solution over source $J$ as 
\begin{equation}
\frac{\delta^k\phi_{cl}(t_1,\vec x_1)}{\delta J^k(t_2,\vec x_2)} = \Phi_k(t_1,\vec x_1; t_2,\vec x_2).
\end{equation}
Then 
\begin{multline}
 \frac{\delta^3 F(\phi_{cl}(t_1,\vec x_1))}{\delta J^3(t_2,\vec x_2)} = \frac{\partial F}{\partial \phi_{cl}}\Phi_3(t_1,\vec x_1; t_2,\vec x_2)\\
 + 3 \frac{\partial^2 F}{\partial \phi_{cl}^2} \Phi_1(t_1,\vec x_1; t_2,\vec x_2)\Phi_2(t_1,\vec x_1; t_2,\vec x_2) \\
 + \frac{\partial^3 F}{\partial \phi_{cl}^3} \Phi_1^3(t_1,\vec x_1; t_2,\vec x_2).
\end{multline}
 Variations  $\Phi_k(t_1,\vec x_1;t_2,\vec x_2)$ can be found by variation of the classical EoM
%\vskip -0.2cm
\begin{equation}
 \frac{\delta^3}{\delta J^3(t_2,\vec x_2)} \left( \partial_{\mu}\partial^{\mu}\phi_{cl}(t_1, \vec x_1) 
 + g^2\phi_{cl}^3(t_1,\vec x_1) = J(t_1, \vec x_1) \right),\nonumber
\end{equation}
 that gives
\begin{multline}\label{variationss}
 L_{t_1}\Phi_1(t_1,\vec x_1;t_2,\vec x_2) = \delta(t_1-t_2)\delta^{(3)}(\vec x_1 -\vec x_2)\\
 L_{t_1}\Phi_2(t_1,\vec x_1;t_2,\vec x_2) = -6 g^2 \phi_{cl}(t_1,\vec x_1)\Phi_1^2(t_1,\vec x_1;t_2,\vec x_2)\\
 L_{t_1}\Phi_3(t_1,\vec x_1;t_2,\vec x_2) = -6 g^2 \Phi_1^3(t_1,\vec x_1;t_2,\vec x_2)\\
 - 18 g^2 \phi_{cl}(t_1,\vec x_1)\Phi_1(t_1,\vec x_1;t_2,\vec x_2)\Phi_2(t_1,\vec x_1;t_2,\vec x_2)\\
 L_{t_1} = \partial^2_{t_1} - \partial^2_{\vec x_1} + 3 g^2 \phi_{cl}^2(t_1,\vec x_1).
\end{multline}
With help of differential equations (\ref{variationss}) we are able to calculate NLO correction to any observable without knowledge 
of exact dependence of the classical solution $\phi_{cl}$ of auxiliary source $J(t,\vec x)$. Although the recipe for NLO calculation 
is fulfilled we can not use it directly with Lagrangian (\ref{static_lagr}).
Analytical solution of equation of motion even for zero source can not be found. 

\section{Spatially homogenous analytical solution }
Let us  consider  $\varphi^4$ toy model . We suppose that all spatial gradients 
small enough to be neglected.
Although this assumption takes us away from implementation to real physical systems
it let us  perform a lot of work analytically and reveal general features that can be 
extended to full theory.

 In spatially homogeneous case $\partial_i \varphi(t, {\bf x}) = 0$ , than 
  \begin{gather}\label{static_lagr_0}
   S = V_3 \int dt\left( \frac{1}{2}\dot \varphi^2-\frac{ g^2}{4}\varphi^4 + J\varphi\right),\\
   V_3 = \int d^3x\nonumber.
  \end{gather}

The equation of motion  is
\begin{gather}
 \ddot \varphi + g^2 \varphi^3 = J.
 \label{EoM}
\end{gather}

We can find classical solution of EoM (\ref{EoM}) for J=0 in terms of the Jacobi elliptical function $cn$ with module \mbox{$k^2 = \frac{1}{2}$ }as
\begin{gather}
 \phi_{cl}(t)=\phi_m cn\left(\frac{1}{2}, g\phi_m t + C\right).
 \label{class_sol}
\end{gather} 
The period of this function is $T_{cl}=\frac{4}{g \phi_m}K(1/2)$, where K(1/2) is the complete elliptic integral of the first kind.
Here $\phi_m$ and $C$ are the amplitude and the phase of the solution.

First of all we calculate  the field expectation value $\langle \hat \varphi \rangle_{t_1}$ as a simplest example.  
In our spatially homogeneous case the eq.(\ref{static_NLO}) has the form of
\begin{gather}\label{phi_NLO_1}
 \langle \hat\varphi\rangle_{t_1}^{LO+NLO} = 
  \Bigg\langle \phi_{cl}(t_1)
  + \frac{g^2 }{4 V_3^2} \int\limits_{t_0}^{t_1}dt_2 
  \phi_{cl}(t_2)\Phi_3(t_1,t_2)  \Bigg\rangle_{i.c.},
\end{gather}
where $\phi_{cl}$ is the Jacobi elliptical function (\ref{class_sol}) and $V_3$ is the full volume of the system (\ref{static_lagr_0}).

Spatially homogeneous case of variations  eq.(\ref{variationss}) depends only on two times. Since we know analytical solution for $\phi_{cl}$ 
and the first variation $\Phi_1(t_1,t_2)$ (see \cite{LR1}) it is clear that 
all three variations are periodic functions (with period of the classical solution (\ref{class_sol})) multiplied by some power of $t_1$. 
It is easy to show that $k$-th variation grows as  $t_1^k$ term at most.

The limits of integration over $t_2$ in NLO term of eq.(\ref{phi_NLO_1}) show that this term is subleading at small times $t_1$ and we can consider
behavior of NLO term only at $t_1>>t_0$. At this limit we can neglect phase $C$ in eq.(\ref{class_sol})  and introduce new dimensionless variable 
$z = g \phi_m t $. It help us obtain  dimensionless variations $f_n(z_1,z_2)$ as 
\begin{multline}
 \phi_{cl}(t_1) = \phi_m f_0(z_1),\\
 \Phi_k(t_1,t_2) = g^{-k} \phi_m^{1-2k} f_k(z_1,z_2), \\ k=1,2,3.
 \label{dimensionless}
\end{multline}
Dimensionless case of eq.(\ref{phi_NLO_1})  has the form
\begin{multline}
 \langle \hat\varphi(t) \rangle_{t_1}^{LO+NLO} = \Big\langle \phi_m f_0(z_1)\\
 \left. +\frac{1}{4 g^2 V_3^2\phi_m^5}\int\limits_{z_0}^{z_1}dz_2 f_0(z_2)f_3(z_1,z_2)
 \right\rangle_{i.c.}.
\end{multline}
Because of periodicity of $f_k(z_1,z_2)$ the integral over $z_2$ behave at late times ($z_1 >> z_0$) as
\begin{gather}
 \int\limits_{z_0}^{z_1}dz_2 f_0(z_2)f_3(z_1,z_2) \approx z_1^3 \psi(z_1), \label{dim_int}
\end{gather}
where $ \psi(z_1)$ is a periodic function with same period $T_{cl}$.
Finally, we have
\begin{multline}\label{phi_NLO_3}
 \langle \hat\varphi(t) \rangle_{t_1}^{LO+NLO} \approx \bigg\langle \phi_m  \bigg[ f_0(g \phi_m t_1)  \\
 +\frac{g t_1^3}{4  V_3^2\phi_m^3} \psi(g\phi_mt_1)\bigg]
 \bigg\rangle_{i.c.} .
\end{multline}
It seems that  NLO term of eq.(\ref{phi_NLO_3}) is dominated at large times $t_1$ before averaging on initial conditions.
Therefore the choice of  the Wigner function play a crucial role and dynamics of the system strongly depends on the initial state. 
This feature is rather general for far-from-equilibrium systems. 

As it is shown in \cite{LR1} in case of gaussian choice for the Wigner function
\begin{gather}\label{Wigner}
 f_W(\alpha,p) =\frac{1}{\pi \alpha_0 p_0} e^{-\frac{(\alpha-A)^2}{\alpha_0^2}}e^{-\frac{p^2}{p_0^2}},\\
 \langle \qquad \rangle_{i.c.} \equiv \int\frac{dp}{2\pi}\int d\alpha f_W(t_0,\alpha,p),\nonumber
\end{gather}
it is possible to perform the averaging over initial conditions analytically. 
 Let us make a change of variables $(\alpha,p) \to (\phi_m,C)$ (see (\ref{class_sol})):
\begin{gather}
\int \frac{d p}{2\pi}  \int d\alpha \to \int |J|\ d \phi_m \ d C.\nonumber\\
|J(\phi_m)|=g\phi_m^2.
\end{gather}
Analytical integration over $\phi_m$ and $C$ is possible in the saddle point approximation, where
\begin{gather}
 f_W(\phi_m,C,0) \approx \frac{1}{\alpha_0 p_0 \pi} e^{-\frac{(\phi_m -A)^2}{\alpha_0^2}
-\frac{C^2 A^4 g^2 }{ p_0^2}}.
\end{gather}
Let us introduce Fourier transform as
\begin{gather}\label{um}
 cn\left(\frac{1}{2}; g\phi_mt + C\right)
 =\sum_{k=-\infty}^{\infty} u_k e^{\frac{2 \pi i k}{T} \left(g\phi_m t +C\right)},\\
 u_m=\frac{1}{T} \int\limits_0^T cn\left(\frac{1}{2};t\right)e^{-i m t\frac{2 \pi}{T}} dt,\nonumber\\
  \psi_m=\frac{1}{T} \int\limits_0^T \psi(t)e^{-i m t\frac{2 \pi}{T}},
\end{gather}
where $T$ is the period of the Jacobi elliptical function $cn\left(\frac{1}{2};t\right)$.
Performing  these calculations  for the field expectation value we obtain
\begin{multline}\label{phi_NLO_2}
 \langle \hat\varphi(t) \rangle_{t_1}^{LO+NLO} \approx 2 A \sum_{k=0}^{\infty}\left( u_k + \frac{g t_1^3}{4  V_3^2 A^3}\psi_k \right) \\
 \times e^{-\frac{ \pi^2 p_0^2}{g^2 A^4 T^2}k^2}
  e^{-\frac{\alpha_0^2 \pi^2 g^2 }{ T^2} k^2 t_1^2}
  cos\left( \frac{2 A g \pi k  }{T} t_1 \right).
\end{multline}
The common time behavior of both LO and NLO terms is oscillation with the amplitude that fall in time according to the cosine and the exponent
of the eq.(\ref{phi_NLO_2}). However NLO term additionally grows as $t_1^3$. The asymptotic behavior of both terms are identical because of  
term $\exp (-t_1^2)$, but early time evolution (usually most interesting part of the problem) fully depend on initial state. 

Parameter $A$ in the Wigner function (\ref{Wigner}) denotes the measure of excitation of the initial field configurations.
It means that starting points of all classical trajectories  $\phi_{cl}(t_0) \equiv \alpha $ are grouped around some value $A$.
In order to describe highly-excited fields at the initial state (that is what we want for heavy ion collisions for example) 
this parameter should be large $A>>1$.
As one can see from (\ref{phi_NLO_2}) it means that NLO term is suppressed.

Similar analysis can be performed for the $T^{\mu\nu}$ components
\begin{gather}\label{Tmn}
 T^{\mu\nu}=\partial^{\mu}\varphi\partial^{\nu}\varphi-
  g^{\mu\nu}\left(\frac{1}{2}\partial_{\sigma}\varphi\partial^{\sigma}\varphi - 
  \frac{g^2}{4}\varphi^4\right).
\end{gather}
At the classical level energy and pressure are:
\begin{gather}
 \varepsilon_{cl} = \frac{1}{2}\dot\phi_{cl}^2+\frac{g^2}{4} \phi_{cl}^4,\ 
 p_{cl} = \frac{1}{2}\dot\phi_{cl}^2-\frac{g^2}{4} \phi_{cl}^4.
\end{gather}
At the LO level we have
\begin{gather}
 \varepsilon_{LO}(t_1) =
 \left\langle \frac{1}{2}\dot\phi_{cl}^2(t_1)+\frac{g^2}{4} \phi_{cl}^4(t_1)
 \right\rangle_{i.c.} ,\\
 p_{LO}(t_1)= \left\langle \frac{1}{2}\dot\phi_{cl}^2(t_1)-\frac{g^2}{4} \phi_{cl}^4(t_1)
 \right\rangle_{i.c.} .
\end{gather}
These result had been obtained by \cite{initial1}. However, one need to note that 
the NLO answers of \cite{initial1} correspond to LO observables in our work.

At the NLO level $\varepsilon_{NLO} = 0$. It is conserved, and  at the initial moment  $\varepsilon_{NLO}(t_0) = 0$
due to integration over $t_2$ in (\ref{static_NLO}). For $p_{NLO}$ we have
\begin{multline}
 \langle \hat p(t) \rangle_{t_1}^{LO+NLO}=\\
 \left\langle g^2\phi_m^4\left[   \pi_{LO}(g \phi_m t_1) + \frac{g t_1^3}{4 V_3^2\phi_m^3} \pi_{NLO}(g\phi_mt_1) \right]
 \right\rangle_{i.c.},\label{p_NLO}
\end{multline}
where 
\begin{gather}
 \pi_{LO}(z_1) = \frac{1}{2}(f_0'(z_1))^2 - \frac{1}{4}f_0^4(z_1)
\end{gather}
 and 
 \begin{multline}
 \int\limits_{z_0}^{z_1}dz_2 f_0(z_1)\Big[ f_0'(z_1)f_3'(z_1,z_2)+3f_1'(z_1,z_2)f_2'(z_1,z_2)-  \nonumber\\
 - f_0^3(z_1)f_3(z_1,z_2)-9 f_0^2(z_1)f_1(z_1,z_2)f_2(z_1,z_2)\nonumber \\
  -6f_0(z_1)f_1^3(z_1,z_2)\Big]\approx z_1^3  \pi_{NLO}(z_1) .
 \label{nonexp}
\end{multline}
Again $\pi_{LO},\ \pi_{NLO}$ are periodic functions with period $T_{cl}$.

  One can note the dimensionless combination 
\begin{equation}
  c_1(g \phi_m t_1) + \frac{g t_1^3}{4 V_3^2\phi_m^3}  c_2(g\phi_mt_1),
\end{equation}
which in both examples eq.(\ref{phi_NLO_3}) and eq.(\ref{p_NLO}) is similar and corresponds to LO + NLO terms. As $\phi_m = \phi_m(\alpha,p)$ 
depends on initial conditions we can see that integration with the Wigner function which gives small amplitude $\phi_m$ (i.e.
has small height and wide tails) results in case where NLO term became significant.  

\section{Conclusions}

Let us summarize the main results presented in the talk.
\begin{itemize}
 \item  Systematic formalism for calculating corrections to
the semiclassical approximation for observables of
highly excited quantum field in the framework of
Keldysh technique is presented. It is shown that
the results obtained in the leading semiclassical
approximation reproduce those within the Classical Statistical Aproximation (CSA) 
thus establishing equivalence between the two approaches. An
important feature of the CSA approximation is that
it is generically not a small-coupling one.
\item  Generic expressions for Next-to-Leading corrections to the
CSA approximation for the scalar field case were
derived.
\item Analytical expressions for the average field, energy
and pressure of the homogeneous scalar field were
derived. The critical role of the character of initial
conditions for applicability of the CSA approximation was discussed.
\end{itemize}

\begin{acknowledgement}
 AR was supported by the Russian Science Foundation under grant No. 16-32-00168
\end{acknowledgement}

\end{document}